\documentclass{article}
\usepackage{waspaa17,amsmath,graphicx,url,times}
\usepackage{color}
\usepackage[ruled]{algorithm2e}
\usepackage{amsmath}
\usepackage{amsfonts}
\usepackage{amssymb}
\usepackage[labelsep=period]{caption}

\def\argmin{\mathop{\rm arg\,min}}

\newcommand{\D}{\displaystyle}

\newcommand{\TmpCapTitle}[0]{default}
\DeclareCaptionFormat{inc-cap-title}{#1: \textbf{\TmpCapTitle#2}#3\par}
\DeclareCaptionFormat{no-cap-title}{#1\textbf{#2}#3\par}

\newcommand{\cLP}{\boldsymbol{c}_{\rho}}
\newcommand{\BS}[1]{\boldsymbol{#1}}

\title{A linear programming approach to the tracking of partials}

\twoauthors
  {Nicholas Esterer}
    {Schulich School of Music, McGill University, \\
     Music Technology Area \& CIRMMT \\
     555 Sherbrooke Street West \\
     Montreal, Quebec \\
     Canada H3A 1E3 \\
     nicholas.esterer@mail.mcgill.ca}
 {Philippe Depalle}
    {Schulich School of Music, McGill University, \\
     Music Technology Area \& CIRMMT \\
     555 Sherbrooke Street West \\
     Montreal, Quebec \\
     Canada H3A 1E3 \\
     philippe.depalle@music.mcgill.ca}

\begin{document}

\ninept
\maketitle

\begin{sloppy}

\begin{abstract}
    A new approach to the tracking of sinusoidal chirps using linear programming
    is proposed. It is demonstrated that the classical algorithm of
    \cite{mcaulay1986speech} is greedy and exhibits exponential complexity for
    long searches, while approaches based on the Viterbi algorithm exhibit
    factorial complexity \cite{depalle1993tracking} \cite{wolf1989finding}. A
    linear programming (LP) formulation to find the best $L$ paths in a lattice
    is described and its complexity is shown to be less than previous
    approaches. Finally it is demonstrated that the new LP formulation
    outperforms the classical algorithm in the tracking of sinusoidal chirps in
    high levels of noise.
\end{abstract}


\section{Introduction}

Atomic decompositions of audio allow for the discovery of meaningful underlying
structures such as musical notes \cite{gribonval2003harmonic} or sparse
representations \cite{plumbley2010sparse}. A classical structure sought in
decompositions of speech and music signals is the sum-of-sinusoids model: windowed sinusoidal atoms in the
decomposition of sufficient energy and in close proximity in both time and
frequency are considered as \textit{connected}. The progressions of these
connected atoms in time form paths or \textit{partial trajectories}.

Many authors have considered the partial tracking problem, beginning with
\cite{mcaulay1986speech}. Their technique is improved upon
in \cite{lagrange2003enhanced} with the use of linear prediction to improve the
plausibility of partial tracks. Rather than seeking individual paths, in
\cite{depalle1993tracking} the most plausible sequence of connections and
detachments between atoms is determined via an extension of the Viterbi
algorithm proposed in \cite{wolf1989finding}. Improvements to this technique are
made in \cite{kereliuk2008improved} by incorporating the frequency slope into
atom proximity evaluations.

The latter techniques seeking globally optimal sets of paths incur great
computational cost due to the large number of possible solutions. For this reason, in
this paper we propose a linear programming \textit{relaxation} formulation of
the optimal path-set problem based on
an algorithm for the tracking of multiple objects in video
\cite{jiang2007linear}. It will be shown that this algorithm has favourable
asymptotic complexity and performs well on the tracking of chirps in high levels
of noise.

\subsection{Note on notation}

The atomic decompositions used in this paper consider blocks of 
of contiguous samples, called \textit{frames} and these frames are
computed every $H$ samples, $H$ being the \textit{hop-size}. We will denote the
$N_{k}$ sets of parameters for atoms found in the decomposition in frame $k$ as $\theta_0^{k},
\dotsc, \theta_{N_{k}-1}^{k}$ and the $N_{k+1}$ in frame $k+1$ as
$\theta_0^{k+1}, \dotsc, \theta_{N_{k+1}-1}^{k+1}$ where $k$ and $k+1$ refer to
adjacent frames.

The total number of nodes is $M = \sum_{k=0}^{K-1} N_{k}$. $\theta_{i}^{j}$ is the
$i$th node of the $j$th frame, $\theta^{j}$ the set of all the nodes in the
$j$th frame and $\theta_{m}$ the $m$th node out of all $M$ nodes ($0 \leq m <
M$).

We are interested in paths that extend across $K$ frames where
each path touches only one parameter set and each parameter set is either
exclusive to a single path or is not on a path.

In this paper, indexing starts at 0. If we have a vector $\BS{x}$ then
$\BS{x}_{i}$ is the $i$th row or column of that vector depending on the
orientation. The same notation is used for Cartesian products, e.g., if $\alpha$
and $\beta$ are sets and $A =
\alpha \times \beta$ then for the pair $a \in A$ $a_{0}$ is the first item in
the pair and $a_{1}$ the second.

\section{A greedy method}

In this section, we present the McAulay-Quatieri method of peak matching. It is
conceptually simple and a set of short paths can be computed quickly, but it can
be sensitive to spurious peaks and its complexity becomes unwieldly for long
searches.

In \cite[p.~748]{mcaulay1986speech} the peak matching algorithm is described in
a number of steps; we summarize them here in a way comparable with the linear
programming formulation to be presented shortly. In that paper, the parameters
of each data point are the instantaneous amplitude, phase, and frequency but
here we allow for arbitrary parameter sets $\theta$.  Define a distance function
$\mathcal{D} \left( \theta_{i},\theta_{j} \right)$ that computes the similarity
between $2$ sets of parameters. We will now consider a method that finds $L$
tuples of parameters that are closest.

We compute the cost tensor
$
    \BS{C} = \theta^{k}
    \otimes_{\mathcal{D}} \mathellipsis \otimes_{\mathcal{D}} \theta^{k+K-1}
$.
For each $l \in \left[0 \dotsc L-1 \right]$, find the indices
$i_{0},\dotsc,i_{K-1}$ 
corresponding to the shortest
distance, then remove the $i_{0},\dotsc,i_{K-1}$th rows (lines of table entries)
in the their respective dimensions from consideration
and continue until $L$ tuples have been determined or a distance between a pair
of nodes on the path
exceeds some
threshold $\Delta_{\text{MQ}}$. This is summarized in
Algorithm~\ref{alg:mq_peak_match}.

\begin{algorithm}
    \KwIn{the cost matrix $\BS{C}$}
    \KwOut{$L$ tuples of indices $\Gamma$, or fewer if $\Delta_{\text{MQ}}$ exceeded}
    $\Gamma \leftarrow \varnothing$\;
    \For{$l \leftarrow 0$ to $L-1$}{
        $\D \Gamma_{l}=\argmin_{[0,\dotsc,M_{0}-1] \times
        \mathellipsis \times [0,\dotsc,M_{K-1}-1] \setminus \Gamma}
            \BS{C}$\;
            \If{$ \exists i,j \in \Gamma_{l} : 
            \mathcal{D} \left( \theta_{i},\theta_{j} \right) > \Delta_{\text{MQ}}$}{
            \KwRet{$\Gamma$}
        }
        $\Gamma \leftarrow \Gamma \cup C_{\Gamma_{l}}$\;
    }
    \KwRet{$\Gamma$}
    \caption{A generalized McAulay-Quatieri peak-matching algorithm.}%
    \label{alg:mq_peak_match}
\end{algorithm}

This is a greedy algorithm because on every iteration the smallest cost is
identified and its indices are removed from consideration. Perhaps choosing a
slightly higher cost in one iteration would allow smaller costs to be chosen in
successive iterations. This algorithm does not allow for that. In other terms,
the algorithm does not find a set of pairs that represent a globally minimal sum of
costs.
Furthermore, the algorithm does not scale well: assuming equal numbers of
parameter sets in all frames, the search space grows exponentially with
$K$. Nevertheless, the method is simple to implement, computationally negligible
when $K$ is small, and works well with a variety of audio signals such as speech
\cite{mcaulay1986speech} and music
\cite{smith1987parshl}.

\section{$L$ best paths through a lattice via linear programming (LP)}

In this section we show how to find L paths through a lattice of $K$ frames such
that the sets of nodes on each path are disjoint. The $k$th frame of the lattice
contains $N_{k}$ nodes for a total of $M = \sum_{k=0}^{K-1}N_{k}$ nodes.

Similar to the McAulay-Quatieri method 
we define the cost $\Delta_{\text{LP}}$ as the limiting cost under which the
connection between two nodes will be considered in the LP method.

The solution vector $\BS{x}$ to the linear program shall indicate the presence of a
connection between a pair of nodes by having an entry equal to $1$ and otherwise
have entries equal to $0$. To enumerate the set of possible connection-pairs we
define
\begin{equation}
    \rho = \left\{ (i,j) : \mathcal{D}(\theta_{i},\theta_{j}) \leq
    \Delta_{\text{LP}} , 0 \leq i < M, 0 \leq j < M, i \neq j \right\}
\end{equation}
The cost vector of the objective function is then 
\begin{equation}
    \cLP = \left\{ D(\theta_{i},\theta_{j}) \forall (i,j) \in
    \rho \right\}
\end{equation}
and the length of $\cLP$ is $\# \rho = \# \cLP = P$, in other words, $P$
pairs of nodes. For convenience we define a bijective mapping ${\mathcal{B} :
\rho \rightarrow [0, \mathellipsis, M-1]}$ giving the index in $\BS{x}$ of the
pair $p \in \rho$. For the implementation considered in this paper,
$\mathcal{D}(\theta_{i},\theta_{j}) = \infty$ for all $i,j$ not in adjacent
frames and so $P$ will be no larger than $(K-1)N^{2}$ (assuming the same
number of nodes $N$ in each frame).

The total cost of the paths in the solution is then calculated through the
inner product $\cLP^{T}\BS{x}$. To obtain $\BS{x}^{\ast}$ that represents $L$
disjoint paths we must place constraints on the structure of the solution. Some
of the constraints presented in the following are redundant but the redundancies
are kept for clarity; later we will show which constraints can be removed
without changing the optimal solution $\BS{x}^{\ast}$.

All nodes in $\BS{x}^{\ast}$ will have at most one incoming connection or
otherwise no connections, a constraint that can be enforced through the
following linear inequality: define $\BS{A}^{\text{I}} \in
\mathbb{R}^{R_{\text{I}} \times P}$ with $R_{\text{I}} = \sum_{k=1}^{K-1}
N_{k}$, the number of nodes in all the frames excluding the first. We sum all
the connections into the node $r_{\text{I}} + N_{0}$ represented by the
respective entry in $\BS{x}$ through an inner product with the $r_{\text{I}}$th
row in $\BS{A}^{\text{I}}$ and require that this sum be between $0$ and $1$,
i.e.,
\begin{equation}
    \BS{A}^{\text{I}}_{r_{\text{I}},\mathcal{B}(p)} = \begin{cases}
        1 & \text{if } p_{1} = r_{\text{I}}+N_{0} \\
        0 & \text{otherwise}
    \end{cases}, 0 \leq r_{\text{I}} < R_{\text{I}}, p \in \rho
\end{equation}
and
\begin{equation}
    \BS{0} \leq \BS{A}^{\text{I}}\BS{x} \leq \BS{1}
\end{equation}
Similarly, to constrain the number of outgoing connections into each node, we
define $R_{\text{O}} = \sum_{k=0}^{K-2} N_{k}$ and
$\BS{A}^{\text{O}} \in \mathbb{R}^{R_{\text{O}} \times P}$ with
\begin{equation}
    \BS{A}^{\text{O}}_{r_{\text{O}},\mathcal{B}(p)} = \begin{cases}
        1 & \text{if } p_{0} = r_{\text{O}} \\
        0 & \text{otherwise}
    \end{cases}, 0 \leq r_{\text{O}} < R_{\text{O}}, p \in \rho
\end{equation}
and
\begin{equation}
    \BS{0} \leq \BS{A}^{\text{O}}\BS{x} \leq \BS{1}
\end{equation}

To forbid breaks in the paths it is required that the number of incoming
connections into a given node equal the number of outgoing connections for the
$R_{\text{B}} = \sum_{k=1}^{K-2} N_{k}$ nodes potentially having both incoming
and outgoing connections.
\begin{equation}
    \BS{A}^{\text{B}}_{r_{\text{B}}} = \BS{A}^{\text{B}}_{r_{\text{B}}} -
    \BS{A}^{\text{B}}_{r_{\text{B}}+N_{0}} \text{ for rows } 0 \leq r_{\text{B}}
    < R_{\text{B}}
\end{equation}
and
\begin{equation}
    \label{eq:cxnbalcon}
    \BS{A}^{\text{B}}\BS{x} = \BS{0}
\end{equation}

Finally we ensure that there are $L$ paths by counting the number of connections
in each frame and constraining this sum to be $L$. We choose arbitrarily to
count the number of outgoing connections by summing rows of $\BS{A}^{\text{O}}$
into rows of $\BS{A}^{\text{C}} \in \mathbb{R}^{(K-1) \times P}$
\begin{equation}
    \BS{A}^{\text{C}}_{r_{\text{C}}} = \sum_{k=a}^{b} \BS{A}^{\text{O}}_{k}
\end{equation}
with $a = \sum_{j=0}^{r_{\text{C}}} N_{j}$ and $b = \sum_{j=0}^{r_{\text{C}}+1}
N_{j}$ and
\begin{equation}
    \label{eq:cxnlcon}
    \BS{A}^{\text{C}}\BS{x} = L\BS{1}
\end{equation}

As stated above, some of these constraints are redundant and can be removed.
Indeed, we have $\BS{0} \leq \BS{x} \leq \BS{1}$, therefore we will always have
$\BS{A}^{\text{I}}\BS{x} \geq 0$ and $\BS{A}^{\text{O}}\BS{x} \geq 0$.
Furthermore, all but the last row of (\ref{eq:cxnlcon}) can be seen as
constructed from linear combinations of rows of (\ref{eq:cxnbalcon}) and the last
row of (\ref{eq:cxnlcon}) so we only require $\BS{A}^{\text{C}}_{K-2}\BS{x} = L$.
Finally we always have $\BS{x} \leq \BS{1}$ because of the constraint that there
be a maximum of $1$ incoming and outgoing connection from each node.

The complete LP to find the $L$ best disjoint paths through a lattice described
by node connections $\rho$ is then
\[
    \min_{\BS{x}} \cLP^{T} \BS{x} 
\]
subject to
\[
    \BS{G}\BS{x} =
    \begin{bmatrix}
        \BS{A}^{\text{I}} \\
        \BS{A}^{\text{O}} \\
        -\BS{I}
    \end{bmatrix} \BS{x} \leq
    \begin{bmatrix}
        \BS{1} \\
        \BS{1} \\
        \BS{0}
    \end{bmatrix}
\]
\begin{equation}
    \label{eq:lpprogfull}
    \BS{A}\BS{x} =
    \begin{bmatrix}
        \BS{A}^{\text{B}} \\
        \BS{A}^{\text{C}}_{K-2}
    \end{bmatrix} \BS{x} = 
    \begin{bmatrix}
        \BS{0} \\
        L
    \end{bmatrix}
\end{equation}
where $\BS{I}$ is the identity matrix. A proof that the solution
$\boldsymbol{x}^{\ast}$ will have entries equal to either $0$ or $1$ can be
found in \cite[p.~167]{parker1988discrete}.  

\section{Memory complexity}

To simplify notation, in this section we assume there are $N$ nodes in each
frame of the lattice.

Although the matrices involved in (\ref{eq:lpprogfull}) are large, only a small
fraction of their values are non-zero. Matrices $\BS{A}^{\text{I}},
\BS{A}^{\text{O}} \in \mathbb{R}^{N(K-1) \times P}$, but each contains only $P$
non-zero entries.  Furthermore $\BS{A}^{\text{B}} \in \mathbb{R}^{N(K-1) \times
P}$ but contains only $2N^{2}(K-2)$ non-zero entries while
$\BS{A}^{\text{C}}_{K-2} \in \mathbb{R}^{P}$ contains merely $N$. The $\BS{x}
\geq \BS{0}$ constraint requires a matrix with $P$ non-zero entries. The total
memory complexity including the entries in $\cLP$ and the right-hand-sides of
(\ref{eq:lpprogfull}) is $2N^{2}(K-2) + 4P + 2N(K-1) + N + 1$ non-zero
floating-point numbers.

\section{Complexity}

Here we will compare the complexity of the LP formulation of the best $L$ paths
search to the greedy McAulay-Quatieri method as well a combinatorial algorithm
proposed in \cite{wolf1989finding}.

Assuming the same number of nodes $N$ in each frame of the lattice, the 
search for the $l$th best path in the generalized McAulay-Quatieri
algorithm ($0 \leq l < L$) requires a search over $(N-l)^{K}$ possible paths.

The LP formulation of the $L$-best paths problem gives results equivalent to the
solution to the $L$-best paths problem proposed in \cite{wolf1989finding}. The
complexity of the algorithm by Wolf in \cite{wolf1989finding} is equivalent to
the Viterbi algorithm for finding the single best path through a trellis whose
$k$th frame has $\binom{N_{k}}{L}\binom{N_{k+1}}{L}L!$ connections where $N_{k}$
and $N_{k+1}$ are the number of nodes in two consecutive frames of the original
lattice. Therefore, assuming a constant number $N$ of nodes in each frame, its
complexity is $O((\binom{N}{L}^{2}L!)^{2}K)$.

The complexity of the algorithm presented here is polynomial in the number of
variables (the size of $\BS{x}$).  Assuming we use the algorithm in
\cite{karmarkar1984new} to solve the LP, our program has a complexity of
$O(P^{3.5}B^{2})$ where $B$ is the number of bits used to represent each number
in the input.  However, this bound is conservative considering the reported
complexity of modern algorithms.

For instance, the complexity of a log-barrier interior-point method is dominated
by solving the system of equations
\begin{equation}
    \label{eq:kkt}
    \begin{bmatrix}
        -\BS{D}\BS{G}^{T}\BS{G} & \BS{A}^{T} \\
        \BS{A} & \BS{0}
    \end{bmatrix}
    \begin{bmatrix}
        \BS{u} \\
        \BS{v}
    \end{bmatrix}
    =
    \begin{bmatrix}
        t\cLP + \BS{A}^{T}\BS{d} \\
        \BS{0}
    \end{bmatrix}
\end{equation}
some $10$s of times \cite[p.~590]{boyd2004convex}. 
Each iteration then takes ${\frac{2}{3}((K-1)N^{2} + (K-2)N)^{3}}$ flops (floating-point
operations) to solve ($\ref{eq:kkt}$) using a standard $LU$-decomposition
\cite[p.~98]{golub1996matrix}. As
$\BS{D}$ is a diagonal matrix, if
the number of nodes in each frame is $N$ for all frames, then
$\BS{D}\BS{G}^{T}\BS{G}$ will be a block-diagonal matrix made up of $K-1$ blocks
$\BS{B}_k \in \mathbb{R}^{N^{2}\times N^{2}}$. The system can then be solved in
\begin{multline*}
    \frac{2}{3}(K-1)N^{6} + 2(K-2)(K-1)N^{5} + \\
    2(K-2)^{2}(K-1)N^{4} + \frac{2}{3}(K-2)^{3}N^{3}
\end{multline*}
flops \cite[p.~675]{boyd2004convex}; this complexity is without exploiting the
sparsity of $\BS{A}$ nor the structure of $\BS{B}_k = \BS{D}_{k}\BS{C}$ --- the product
of some diagonal matrix $\BS{D}_{k}$ with an unchanging symmetric matrix
$\BS{C}$.

\section{Partial paths on an example signal\label{sec:mq_lp_compare_chirp}}

We compare the greedy and LP based methods for peak matching on a synthetic
signal. The signal is composed of $Q=3$ chirps of constant amplitude, the $q$th
chirp $s$ at sample $n$ described by the equation
\[
    s_{q}(n) = \exp(j(\phi_{q} + \omega_{q}n +
    \frac{1}{2} \psi_{q} n^{2}))
\]
The parameters for the $Q$ chirps are presented in
Table~\ref{tab:ptrackexamplechirpparams}.

\begin{table}[!b]
    \caption{Parameters of $q$th chirp. $\nu_{0}$ and $\nu_{1}$ are the initial and
    final frequency of the chirp in Hz. \label{tab:ptrackexamplechirpparams}}
    \begin{center}
        \begin{tabular}{l c c c c c}
            $q$ & $\phi_{q}$ & $\omega_{q}$ & $\psi_{q}$ & $\nu_{0}$ & $\nu_{1}$ \\
            \hline
            0 & 0 & 0.20 & 2.45 $\times 10^{-6}$ & 500 & 600 \\
1 & 0 & 0.39 & 4.91 $\times 10^{-6}$ & 1000 & 1200 \\
2 & 0 & 0.59 & 7.36 $\times 10^{-6}$ & 1500 & 1800 \\

        \end{tabular}
    \end{center}
\end{table}

A 1 second long signal is synthesized at a sampling rate of 16000 Hz, the
chirps ramping from their initial to final frequency in that time. We add
Gaussian distributed white noise at several SNRs to evaluate the technique in the
presence of noise.

\begin{figure*}[!t]
    \centering
    \centerline{\includegraphics[width=0.75\textwidth]{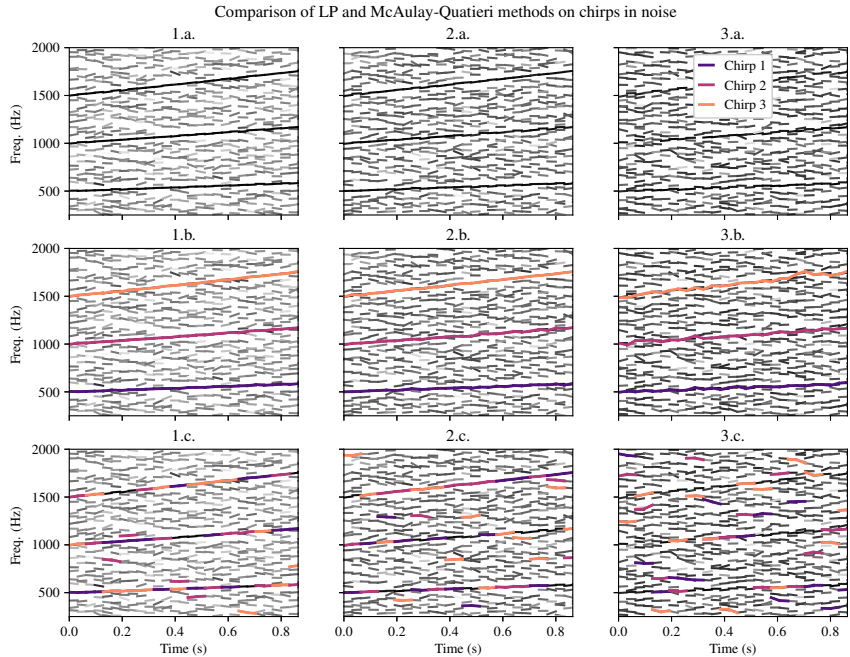}}
    \caption{ Line-segments representing the frequency and frequency-slope at
        local spectrogram maxima. The power of each atom is represented by
        shades of grey: black atoms have the highest power and white the lowest.
        The coloured segments correspond to connected paths returned by the
        search algorithms. Plots 1--3.a. show the atomic decomposition of the
        signal before partial tracking for SNRs of 0, -6 and -12 dB,
        respectively.  Plots 1--3.b. show the partial paths discovered by the LP
        method and plots 1--3.c. show the paths discovered by the
        McAulay-Quatieri method.  See Table~\ref{tab:ptrackexamplechirpparams}
        for the chirp parameters.  \label{plot:mq_lp_compare_chirps}
    }
\end{figure*}

A spectrogram of each signal is computed with an analysis window length of 2048
samples and a hop-size $H$ of 512 samples. Local maxima are searched in 100 Hz
wide bands spaced 50 Hz apart. The bin corresponding to each local maximum and its two
surrounding bins are used by the Distribution Derivative Method
(DDM) \cite{betser2009sinusoidal} to estimate the
local chirp parameters, the $i$th set of parameters in frame $k$ denoted
$\theta_{i}^{k} = \left\{ \phi_{i}^{k} , \omega_{i}^{k} , \psi_{i}^{k}
\right\}$ (the atoms used by the DDM are generated from 4-term
once-differentiable Nuttall windows \cite{nuttall1981some}). Partial tracking is performed on the
resulting atomic decomposition.

We search for partial tracks using both the greedy and LP strategies. Both
algorithms use the distance metric $\mathcal{D}_{\text{pr.}}$ between two parameters sets:
\begin{equation}
    \label{eq:examplecostfun}
    \mathcal{D}_{\text{pr.}} \left( \theta_{i}^{k},
    \theta_{j}^{k+1} \right) = \left( \omega_{i}^{k} +
    \psi_{i}^{k} H - \omega_{j}^{k+1} \right)
\end{equation}
which is the error in predicting $j$th frequency in frame $k+1$ from the $i$th
parameters in frame $k$. For the greedy method, the search for partial paths is
restricted to two frames ahead, i.e., paths of length $K_{\text{MQ}}=3$ are sought, otherwise the
computation becomes intractable. For the LP
method the search is carried out over all frames ($K_{\text{LP}}=28$).
The cost thresholding values are ${\Delta_{\text{MQ}} = \Delta_{\text{LP}} =
0.1}$. For both methods, the search is restricted
to nodes between frequencies 250 to 2000 Hz.

Figure~\ref{plot:mq_lp_compare_chirps}
shows discovered partial trajectories for signals at various SNRs. It is seen
that while the greedy method starts performing poorly at an SNR of -6 dB, the LP method
still gives plausible partial trajectories. The LP method returns paths spanning
all $K$ frames, due to the constraints. The McAulay-Quatieri method in general does not,
but longer paths can be formed in a straightforward way after the initial short path
search step \cite{mcaulay1986speech}.

It is interesting to note that the paths are found by only considering the
prediction error of the initial frequency of the atom. Other cost functions can
be chosen depending on the nature of the signal: reasonable cost functions here
might be similarity of the atoms's energies or frequency slopes.

\section{Conclusion}

In this paper we reformulated the classical partial tracking technique of McAulay and
Quatieri and showed that it can be seen as a greedy algorithm for finding the $L$
shortest paths in a lattice. An algorithm was then proposed minimizing the sum
of the $L$ paths, using a linear programming approach. The complexity of the new
algorithm was shown to be generally less than the Viterbi-based methods and
the McAulay-Quatieri algorithm for large $K$.
It was shown on synthetic signals that the new approach finds plausible paths in
lattices with a large number of spurious nodes.

The proposed approach has some drawbacks.
There are situations where it is undesirable to have paths extend
throughout the entire lattice. Acoustic signals produced by striking media, such
as strings or bars, exhibit a spectrum where the upper partials decay more
quickly than the lower ones; it
would be desirable in these situations to have shorter paths for these
partials. This could be addressed as in
\cite{depalle1993tracking} where the signal is divided into overlapping sequences of
frames and partial paths are connected between sequences.

In its current form, the path search may choose undesirable paths if a
convenient node is missing from the following frame. An extension could consider
nodes some number of frames ahead.

The proposed algorithm, while asymptotically faster than other partial tracking
algorithms, is still not fast. In situations where computational resources
are limited, a McAulay-Quatieri method search over many sets of small $K$ works
sufficiently well. However in high amounts of noise the algorithm proposed here
is robust while still of tractable complexity.

It may be possible to improve the performance of the algorithm by the use of
different cost functions and regularization. The cost function
(\ref{eq:examplecostfun}) could be extended to encourage similarity between
frequency slopes or amplitude information. Each metric should be scaled
according to its desired contribution to the cost.

There may also be a way to extract individual paths through the use of auxiliary
variables in (\ref{eq:lpprogfull}). If so, path specific costs such as overall
smoothness or fit to a particular model could be incorporated. 

In any case, it would be interesting to further investigate programming
relaxations encouraging underlying discrete structures plausible for audio
in the framework of regularized approximation. These structures are closer to
ground-truth structures for speech (text) and music (the musical score).


\bibliographystyle{IEEEtran}
\bibliography{paper}

\end{sloppy}
\end{document}